\documentclass[lettersize,journal]{IEEEtran}
\IEEEoverridecommandlockouts
\usepackage{cite}
\usepackage{amsmath,amssymb,amsfonts}
\usepackage{algorithmic}
\usepackage{enumitem}
\usepackage{subfigure}
\usepackage{graphicx}
\usepackage{textcomp}
\usepackage{gensymb}
\usepackage{xcolor}
\usepackage{soul}
\usepackage{float}
\usepackage{algorithm}
\usepackage{cite}
\usepackage{soul}
\usepackage{gensymb}
\usepackage{etoolbox}
\usepackage{rotating}
\usepackage{comment}
\usepackage{balance}
\usepackage{booktabs}
\usepackage{multirow}
\usepackage{hyperref}
\usepackage{CJK} 
\usepackage{tikz}
\usepackage{algorithmic}
\usepackage{algorithm}
\usepackage{array}
\usepackage{textcomp}
\usepackage{stfloats}
\usepackage{url}
\usepackage{verbatim}
\usepackage{graphicx}
\usepackage{cite}
\usepackage{wrapfig}
\usepackage{tabularx}
\usepackage{longtable}
\usepackage{float}
\usepackage{soul}
\usepackage{multirow}

\usepackage{enumitem}
\usepackage{booktabs}
\usepackage{hhline}
\usepackage{boldline} 
\usepackage{ragged2e}
\usepackage{mathtools}
\usepackage{breqn}
\usepackage{color}

\newcolumntype{P}[1]{>{\centering\arraybackslash}p{#1}}
\newcolumntype{M}[1]{>{\arraybackslash}m{#1}}

\def\BibTeX{{\rm B\kern-.05em{\sc i\kern-.025em b}\kern-.08em
    T\kern-.1667em\lower.7ex\hbox{E}\kern-.125emX}}
\begin{document}
\bstctlcite{IEEEexample:BSTcontrol}

\title{Free-Space Optical Channel Turbulence Prediction: A Machine Learning Approach}

\author{
Md Zobaer Islam, Ethan Abele, Fahim Ferdous Hossain, Arsalan Ahmad, \IEEEmembership{Senior Member, IEEE}, \\Sabit Ekin, \IEEEmembership{Senior Member, IEEE}, and John F. O'Hara, \IEEEmembership{Senior Member, IEEE}
\thanks{This work was supported in part by the National Aeronautics and Space Administration under Grant 80NSSC20M0214.~(\textit{Corresponding author: Ethan Abele, John F. O'Hara.})}
%\thanks{Md Zobaer Islam is with the Department of Radiology, University of North Carolina at Chapel Hill, Chapel Hill, North Carolina, USA, and the School of Electrical and Computer Engineering, Oklahoma State University, Oklahoma, USA (e-mail: zobaer\_islam@med.unc.edu, zobaer.islam@okstate.edu)
%}
%\thanks{Ethan Abele, Fahim Ferdous Hossain, and John F. O'Hara are with the School of Electrical and Computer Engineering, Oklahoma State University, Oklahoma, USA (e-mail: eabele, fferdou, oharaj\{@okstate.edu\})
%}
\thanks{Md Zobaer Islam, Ethan Abele, Fahim Ferdous Hossain, and John F. O'Hara are with the School of Electrical and Computer Engineering, Oklahoma State University, Oklahoma, USA (e-mail: zobaer.islam, eabele, fferdou, oharaj\{@okstate.edu\})
}
\thanks{Arsalan Ahmad is with the Department of Electrical and Computer Engineering, Iowa State University, Ames, Iowa, USA (e-mail: aahmad@iastate.edu)
}
\thanks{Sabit Ekin is with the Departments of Engineering Technology, and Electrical \& Computer Engineering, Texas A\&M University, College Station, Texas, USA (e-mail: sabitekin@tamu.edu)
}
}

\markboth{IEEE COMMUNICATIONS LETTERS,~Vol.~XX, No.~X, XXX~2025}%
{Shell \MakeLowercase{\textit{et al.}}: A Sample Article Using IEEEtran.cls for IEEE Journals}

\maketitle

%\vspace{-6mm}
\begin{abstract}
Channel turbulence is a formidable obstacle for free-space optical (FSO) communication.  Anticipation of turbulence levels is highly important for mitigating disruptions but has not been demonstrated without dedicated, auxiliary hardware. We show that machine learning (ML) can be applied to raw FSO data streams to rapidly predict channel turbulence levels with no additional sensing hardware.  FSO was conducted through a controlled channel in the lab under six distinct turbulence levels, and the efficacy of using ML to classify turbulence levels was examined.  ML-based turbulence level classification was found to be $>98$\% accurate with multiple ML training parameters.  Classification effectiveness was found to depend on the timescale of changes between turbulence levels but converges when turbulence stabilizes over about a one minute timescale.
\end{abstract}

%\vspace{-3mm}

\section{Introduction}

Free-space optical (FSO) communication is emerging as a critical technology for high-speed, wireless transmission of data in certain applications.  It can bypass terrain where guided-wave communication systems are impractical, and it can be rapidly deployed in disaster areas.  FSO also offers enhanced security, reduced size, weight, and power (SWaP), and increased bandwidth compared to radio frequency (RF) systems. These advantages have driven wide adoption of FSO in applications from satellite communications to terrestrial point-to-point links. 

%\vspace{.5mm}

Nevertheless, FSO still faces significant practical challenges, such as fog, pointing error, and atmospheric turbulence. The last of these is a major source of signal degradation, even in clear weather~\cite{sahu2018,prabu2014}. Turbulence is a phenomenon that refers to the random fluctuations in the refractive index of the atmosphere \cite{andrews2005laser}. These fluctuations cause variations in the intensity and phase of the transmitted optical signal, leading to beam wander and scintillation at the receiver \cite{andrews2006strehl}. The results are signal fading, distortion, increased bit-error rates, and generally worsened performance/reliability.  

%\vspace{.5mm}

\begin{figure*}[ht]
\includegraphics[width=0.90\textwidth]{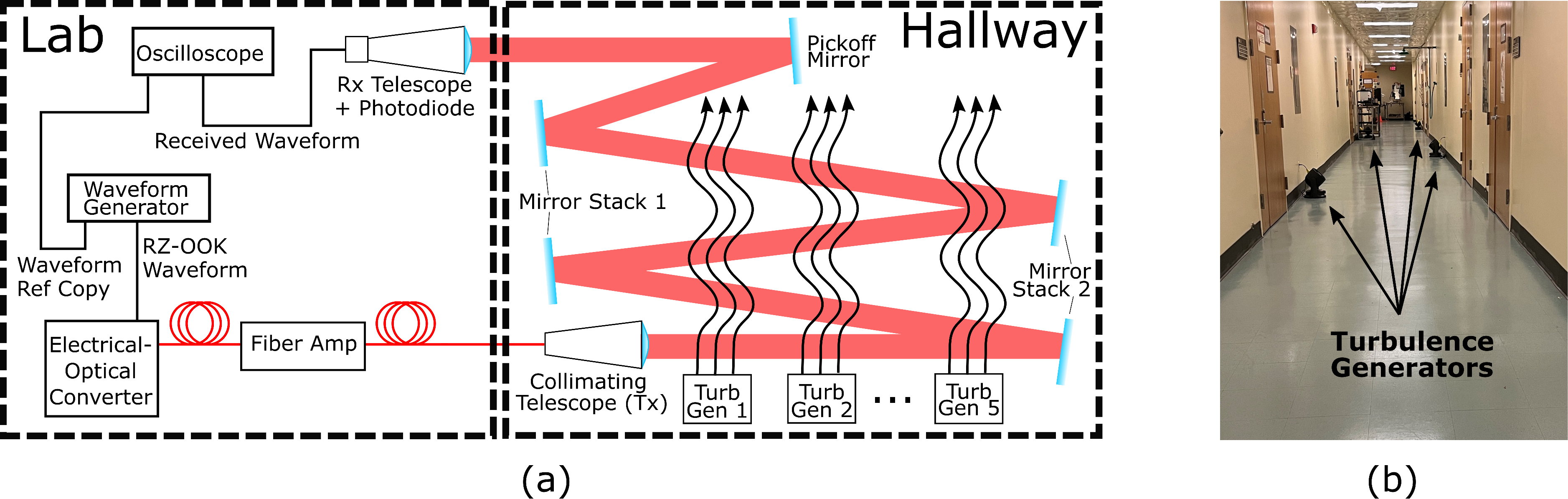} 
\centering
\vspace{-2mm}
\caption{(a) System diagram showing the general connections between instruments and turbulence flow through the optical path, (b) View of the lab hallway with four turbulence generators marked}.
\label{fig:diagram}
\vspace{-4mm}
\end{figure*}

Turbulence is a major loss factor in most FSO systems \cite{majumdar2010free,hemmati2020near}and link budgets rely on accurate turbulence estimation for both Earth-to-space and terrestrial applications.  Much effort has been focused on overcoming it by various means, including increased transmitted power, adaptive optics, and  optimization of beam width \cite{zhao2010optimum}. However, many of these methods require or benefit from an accurate estimate of the turbulence levels.  A common  parameter used in quantifying turbulence level is the refractive index structure constant $C_n^2$, which can vary greatly in position and time \cite{odintsov2019determination,fiorino2021re}. Orders of magnitude difference can be seen from morning to night, with large fluctuations also occurring on the scale of minutes. Several models have been developed to predict $C_n^2$, but their accuracy is limited when used outside the region where they were derived \cite{hemmati2020near}.  Turbulence measurement techniques also exist, including scintillometers, image-based techniques, and balloon soundings \cite{jian2023review}.  However, these  also suffer drawbacks: image-based techniques require a visible target whose distorted image can be monitored, limiting the maximum range. Balloon-based techniques take hours to rise through the atmosphere and complete their readings.  Acoustic sounding systems add significantly to the cost, power requirement, and size of the ground station installation. FSO systems would greatly benefit from methods to quickly classify local or distance-integrated turbulence levels, particularly if existing hardware could be leveraged. This information could improve the efficiency and reliability of FSO systems by enabling adjustment of power and beam parameters, possibly in real time. For hybrid RF/FSO systems, the accurate prediction of turbulence may also underpin informed switching decisions between RF and FSO links, ultimately enhancing the overall system performance and reliability.

Conventional methods for predicting or classifying turbulence rely on complex mathematical models and physical simulations~\cite{puryear2011,kashani2015,lionis2022}. These approaches suffer from limitations, such as impracticality in real-time applications~\cite{quatresooz2023} and dynamic environments.  Machine learning (ML) and deep learning (DL) present a compelling and potentially much simpler alternative for turbulence prediction in FSO channels~\cite{lionis2023experimental,lohani2018turbulence,bakir2023deep,elmabruk2024atmospheric}. By leveraging advanced algorithms, ML and DL models enable the classification of received optical data in a more flexible and practical manner than traditional methods. The ability of these models to discern complex patterns and relationships makes them good candidates for predicting turbulence and facilitating adaptive communication strategies~\cite{esmail2021free}. In this study, we test and affirm that ML-based turbulence prediction is feasible by providing empirical analysis of its application in classifying received FSO data transmitted through various turbulent channels.  We also present details of the ML approach and its practical limitations.

The rest of this paper is organized as follows. Section~\ref{sec:System_Design} describes the experimental details. Section~\ref{sec:data_coll} presents the measured data, their collection scheme, and discussion surrounding the bit error rate (BER) and Q-factor . Section~\ref{sec:turbpred} discusses the ML approach to turbulence  prediction and associated results. Insights drawn from the data and results are discussed in Section~\ref{sec:disc}. Finally, Section~\ref{sec:Conclusion} presents some conclusions and future directions.

\section{System Design and Implementation}
\label{sec:System_Design}
%Describe the system model and experimental setup. \hlgreen{Need help from other authors here. This section should include the details of the experimental setup with a diagram showing its important blocks. Use necessary references of different equipment used as well. Additionally, include relevant photos of the setup.}

A turbulence test chamber was constructed within the hallway of the Ultrafast Terahertz and Optoelectronic Laboratory (UTOL) at Oklahoma State University. The hallway, shown in Fig.~\ref{fig:diagram}, is well suited for testing atmospheric effects. It is sealed by doors at either end, providing relative isolation from activity and airflow in the rest of the building. This allows the turbulence in the chamber to be altered in a controlled manner. 

Figure~\ref{fig:diagram} illustrates the equipment setup. The transmit signal was provided by a Keysight M8195A arbitrary waveform generator (AWG), capable of creating electrical signals with 25\,GHz of bandwidth. For this work, a pseudo-random binary sequence (PRBS) was used to randomize the bit stream and avoid any bias in the results. The AWG output modulated a 1550\,nm electrical-optical converter. The modulated signal was transmitted through the chamber using an optical system with parameters described in Table \ref{table:opticalParams}.

\begin{table}[h]
\renewcommand{\arraystretch}{1.24}
\centering
\caption{Optical System Parameters}
\vspace{-3mm}
\begin{center}
\begin{tabular}{|P{2.9cm}|P{3cm}|P{1cm}|}  
\hline
\centering\bf{Parameter}  & \bf{Value} & \bf{Unit} \\
\hline
\centering Tx Transmit Power  & $\approx 5$ & dBm \\ 
\hline
\centering Tx Aperture Diam.  & 50.8 & mm\\
\hline
\centering Hall Mirrors  & Silver Coated, $R \geq 98\%$ & -\\
\hline
\centering Rx Aperture Diam.  & 75 & mm\\
\hline
\centering Rx Detector  & InGaAs APD & -\\
\hline
\centering Rx Responsivity  & 9 & A/W\\
\hline
\centering Background Noise (est.) & -12 & dBm\\
\hline
\end{tabular}
\end{center}
\label{table:opticalParams}
\vspace{-3mm}
\end{table}

The beam traversed the hallway four times in total before a final pick-off mirror directed the beam out of the hallway to a receive (Rx) telescope, giving a total propagation distance of approximately 172\,m. The electrical signal from the photodetector was captured on a Keysight DSOV254A digital storage oscilloscope (DSO). Turbulence generators were placed approximately 6~m apart along the hallway to direct turbulent airflow across the beam path. The generators were commercial fans with 1500~W heating elements. In our previous work, imaging-based turbulence measurements were performed to independently quantify the extent of achievable turbulence within the test chamber. Kolmogorov theory was then applied to the measurements to show that the system can generate turbulence over a significant range, i.e. $C_n^2$ values from $4.8\times10^{-16}$ (low turbulence) to $5.0\times10^{-14}$ m$^{-2/3}$ (moderate turbulence)\cite{abele2023channel}.

%This allows a variable degradation in received signal power, %which results from random wander of the beam. The variance of %beam displacement is commonly related to $C_n^2$ by %\cite{abele2023channel}:

%\begin{equation}
%\langle\beta^2\rangle = %0.54H^2\sec^2\left(\theta\right)\left(\frac{\lambda}{2W_0}\right)^2\left(\frac{2W_0}{r_0}\right)^{5/3}
%\label{eqn:misalignment-gain}
%\end{equation}

%\begin{equation}
%r_0 = \left[ 0.432\int_{0}^{H}C_n^2(h)dh \times %k^2\sec(\theta)\right]^{-3/5}
%\label{eqn:misalignment-gain}
%\end{equation}

%where H is the propagation distance, $r_0$ is the Fried parameter, $\lambda$ is the carrier wavelength, and the zenith angle $\theta$ may be regarded as zero in our setup. In \cite{abele2023channel} this system was shown to produce $C_n^2$ values from $4.8\times10^{-16}$ to $5.0\times10^{-14}$. 

%\begin{figure}[!htp]
%\includegraphics[width=0.45\textwidth]{HallwayPath.png} 
%\centering
%\caption{Indoor turbulence chamber for experimental measurements.}
%\label{fig:testbed}
%\vspace{-3mm}
%\end{figure}

%\vspace{-3mm}
\section{Data Collection}
\label{sec:data_coll}
Optical communication data was sent through the system in Section~\ref{sec:System_Design} and analyzed under several discrete turbulence levels.  The transmitted PRBS had a length of 2040 bits plus a 32 bit header so that its beginning could be identified in the measured data. Two repetitions of this sequence were transmitted sequentially in return-to-zero, on-off-keying (RZ-OOK) format. The transmitted data rate was 5\,Mbps, which was chosen so that long measurements could be acquired without exceeding the memory depth of the DSO. Measurements on the order of seconds are required to observe turbulence effects. 

The AWG was configured to output the  waveform simultaneously on two output channels. One channel provided a modulation signal to the laser transmitter. The laser traveled through the optical path in the test chamber and was collected by the Rx telescope. The resulting photodiode voltage was recorded on one channel of the DSO. The second AWG output fed directly into another channel of the DSO and served as ground truth (see Fig.~\ref{fig:diagram}). The output signal was continuously repeated so that each capture contained many thousands of bits. 

Turbulence in the test chamber was varied by changing the number of active turbulence generators in the optical path. Turbulence levels ranged from 0 to 5 fans, representing six distinct turbulence levels, with 0 denoting the least and 5 denoting the highest turbulence levels. Based on our prior chamber measurements, these turbulence levels all fell within the aforementioned range of $C_n^2$. The DSO captured the received signal for 2.5\,s in three repetitions, generating three data files for each turbulence level. The incoming waveforms were recorded at 8 samples per bit or 40 MSps. The duration of data collection was constrained by the memory depth of the DSO ($\approx100$ Msamples). Data were captured twice with the above procedure -- once immediately after activating the turbulence generators, and again after a stabilization period of 10~min, during which the generators operated continuously. 

\begin{figure*}[!htp]
\includegraphics[width=0.85\textwidth]{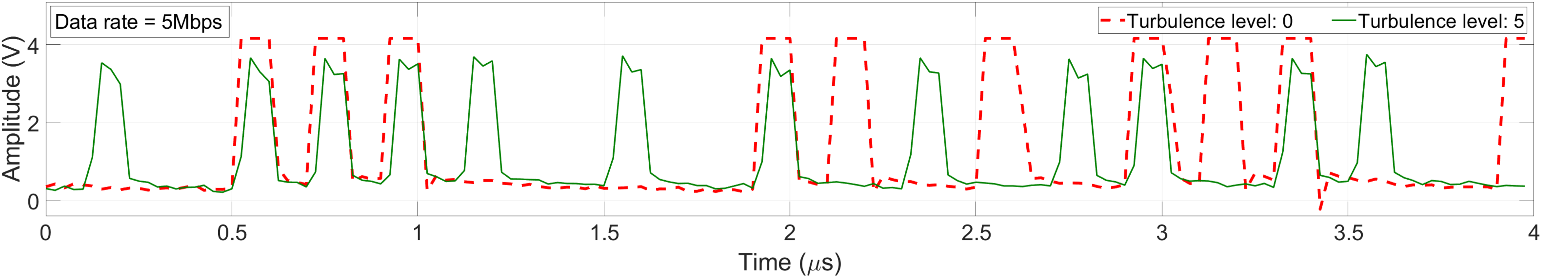} 
\centering
\vspace{-3mm}
\caption{Received data plots from the beginning of the first files of turbulence levels 0 and 5.}
\label{fig:dplot}
\vspace{-4mm}
\end{figure*}

Fig.~\ref{fig:dplot} shows plots of the first 160 samples for the minimum and maximum turbulence levels with zero stabilization time. Thus, it presents the first 20 bits of data. The graphs reveal that the bits exhibit cleaner and flatter tops for `1' bits in the absence of turbulence (level 0). Conversely, at turbulence level 5, the bits exhibit partial distortion due to channel degradation resulting from higher turbulence.  Since `0' bits are represented by zero laser intensity in OOK, their noise variance is unaffected by the turbulence. The BER and Q-factor of the measured signals were generated for each turbulence condition to quantify the effects of the turbulence on communications. The BER was computed by comparing the demodulated optical signal and the reference data across more than 12 million bits. The Q-factor was estimated from the sampled bit values using the procedure described in \cite{abele2023channel}. The results are listed in Table \ref{table:BER_Q}. The measurements show a degradation of Q with increasing turbulence, as expected.  However, a lower transmit power was used in the `10~min delay' measurement which resulted in higher overall BER. Probability density functions (PDFs) for similar data can be found in our prior work \cite{abele2023channel}.

\begin{table}[t]
\renewcommand{\arraystretch}{1.24}
\centering
\caption{Signal Performance Metrics}
\vspace{-6mm}
\begin{center}
\begin{tabular}
{|P{0.7cm}|P{0.5cm}|P{0.7cm}|P{0.7cm}|P{0.7cm}|P{0.7cm}|P{0.7cm}|P{0.7cm}|}
\hline
\multicolumn{2}{|c|}{\multirow{2}{*}{\centering\bf{Delay/Metric}}} & \multicolumn{6}{p{6cm}|}{\centering\bf{Channel Conditions}}\\
\cline{3-8}
\multicolumn{2}{|c|}{} &\bf{0 Fan} & \bf{1 Fan} & \bf{2 Fan} & \bf{3 Fan} & \bf{4 Fan} & \bf{5 Fan}\\
\hline
\multirow{2}{*}{\centering\bf{None}} & \bf{Q} & 39.30 & 24.71 & 10.27 & 7.79 & 8.62 & 7.75 \\
\cline{2-8}
& \bf{BER} & 0 & 0 & 0 & 0 & 0 & 0\\
\hline
\multirow{2}{*}{\centering\bf{10 Min}} & \bf{Q} & 6.36 & 5.94 & 2.02 & 2.07 & 1.71 & 1.75 \\
\cline{2-8}
& \bf{BER} & 0 & 0 & 0.0031 & 0.0032 & 0.0062 & 0.0101 \\
\hline
\end{tabular}
\end{center}
\label{table:BER_Q}
\vspace{-6.0mm}
\end{table}

\section{Turbulence Prediction}
\label{sec:turbpred}

Extreme gradient boosting or XGBoost, an ensemble ML algorithm, was utilized to classify the received data based on the turbulence level. Ensemble learning methods offer enhanced classification accuracy compared to individual models, such as single decision trees, by integrating multiple models in an optimized manner. There are two primary pathways in ensemble learning: bagging and boosting~\cite{sagi2018}. Bagging techniques average the results of multiple independently-built base models. Boosting techniques excel in the bias-variance trade-off compared to bagging techniques by sequentially refining models to correct errors made by preceding ones, thereby focusing on the most informative features of the data~\cite{fatima2023xgboost}. XGBoost is a popular boosting technique due to its superior power and efficiency over traditional methods like AdaBoost or gradient boosting. This superiority arises from its enhanced regularization techniques, efficient feature splitting, and parallel processing capabilities. In this study, XGBoost algorithm was implemented using the Scikit-learn library with default parameters \cite{xgb} in the Python programming language.

%ensemble learning method that uses decision trees as its base model and a serial boosting technique to make its predictions more accurate~\cite{sagi2018}. It builds a strong model by combining many weak models sequentially. Each base model learns from a different subset of the data, and it corrects errors made by the previous ones. This helps XGBoost avoid common problems like overfitting or underfitting and improves its overall prediction accuracy. As it learns, XGBoost gives more weight to the base models that do a better job of reducing errors or improving the model. Thus, it pays attention to the most informative features of the data and those that contribute most significantly to the final prediction.

%\begin{figure}[!htbp]
%\includegraphics[width=0.5\textwidth]{difffiles2.png} 
%\centering
%\caption{Turbulence classifiction accuracies (data collected from beginning of the files, train+test data count=4500, number of bits per data=1000)}
%\label{fig:restemp2}
%\end{figure}

The class labels ranged from 0 to 5 in this study, corresponding to the turbulence level set in the chamber. To diversify the strategy for training and predicting turbulence levels, distinct approaches were employed for selecting training and testing data. To remove bias from the data, they were normalized to a range between 0 and 1 before being fed to the ML model. At first we considered three data files (files 1, then 2, then 3), collected immediately and sequentially after activating the turbulence generators. There was a $\approx10$~second time gap between consecutive files, meaning file 3 was collected within $<1$~min of file 1.  In the initial training phase, the data from the beginning of each file were independently extracted, with 1000 bits per data signal and a total data count of 4500. A random 80\%-20\% train-test split was performed, and the turbulence classification was conducted using the XGBoost algorithm. %Subsequently, various combinations of data from two files were jointly considered for training and testing, while maintaining a consistent total number of bits. Finally, the impact of the number of files considered on classification accuracy was assessed by averaging accuracies for scenarios involving 1 and 2 files, and the accuracy achieved when all three files were considered together.
Next, the same analysis was executed on the dataset collected after a 10~min turbulence stabilization time. Turbulence prediction accuracies on test datasets for both cases are shown in Table~\ref{table:file123}. These accuracies indicate the success rate of the ML algorithm in predicting the channel turbulence level (0-5), calculated as the ratio of the number of correctly classified instances to the total number of instances in the dataset in each case.

Next, another analysis using the data with zero stabilization duration was conducted.  In this case, the number of bits per data instance, or equivalently the time duration considered per data, was varied to see its effect on turbulence prediction accuracy. While keeping the total number of data instances constant at 4500, the number of bits per data was varied from 200 to 3000 for all three files independently for both training and testing purposes.  Again, turbulence classification was conducted with an 80\%-20\% train-test split. The classification accuracy scores on test data resulting from this investigation are presented in Fig.~\ref{fig:varybitsandlength}(a). Additionally, an examination was performed on the effect of varying the number of data instances. This analysis kept the number of bits per data instance constant at 1000, while varying the number of total data instances considered. % within 780 to 33000.
The corresponding prediction accuracy scores on test data are displayed in Fig.~\ref{fig:varybitsandlength}(b). For additional insights and comparisons, similar graphs were generated with the data collected after a 10~min stabilization time, and these are presented in Fig.~\ref{fig:varybitsandlength10mins}. The ranges of $x$-axes in both Fig.~\ref{fig:varybitsandlength} and Fig.~\ref{fig:varybitsandlength10mins} were based on convenient segmentation of data instances from the original longer-duration data files.

\begin{table}[t]
\renewcommand{\arraystretch}{1.24}
\centering
\caption{Turbulence classification test accuracies}
\vspace{-3mm}
\begin{center}
\begin{tabular}{|M{1.9cm}|P{2cm}|P{3.1cm}|} 
\hline
\multirow{2}{*}{\bf{Files considered}} & \multicolumn{2}{p{5.1cm}|}{\centering\bf{Turbulence classification accuracy scores}}\\
\cline{2-3}
& \bf{no stabilization} & \bf{10 minutes stabilization} \\
  \hline
\centering\bf{File 1}  & 90.00\%   & 98.33\%   \\
\hline
\centering\bf{File 2}  & 98.56\%    & 98.11\%  \\
\hline
\centering\bf{File 3}  & 99.11\%   & 99.44\%   \\
  \hline
\end{tabular}
\end{center}
\label{table:file123}
\vspace{-6mm}
\end{table}

\vspace{-3mm}
\section{Discussion}
\label{sec:disc}
\begin{figure}[!htbp]
\includegraphics[width=0.465\textwidth]{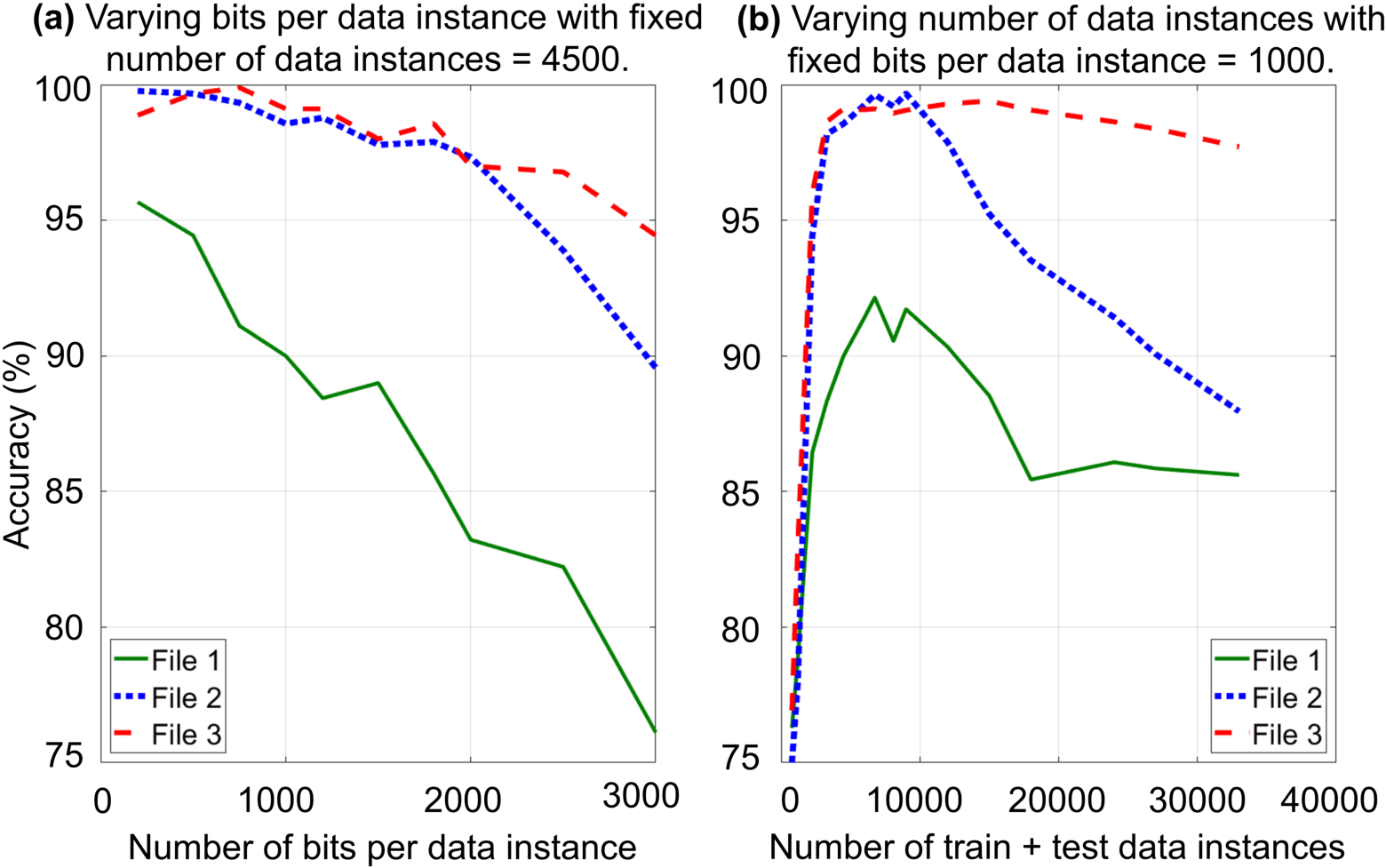} 
\centering
\vspace{-2mm}
\caption{Turbulence classification test accuracy scores for different files with zero stabilization time after activation of the turbulence generators (data were taken from the beginning of each file).}
\label{fig:varybitsandlength}
\vspace{-2mm}
\end{figure}

\begin{figure}[!htbp]
\includegraphics[width=0.465\textwidth]{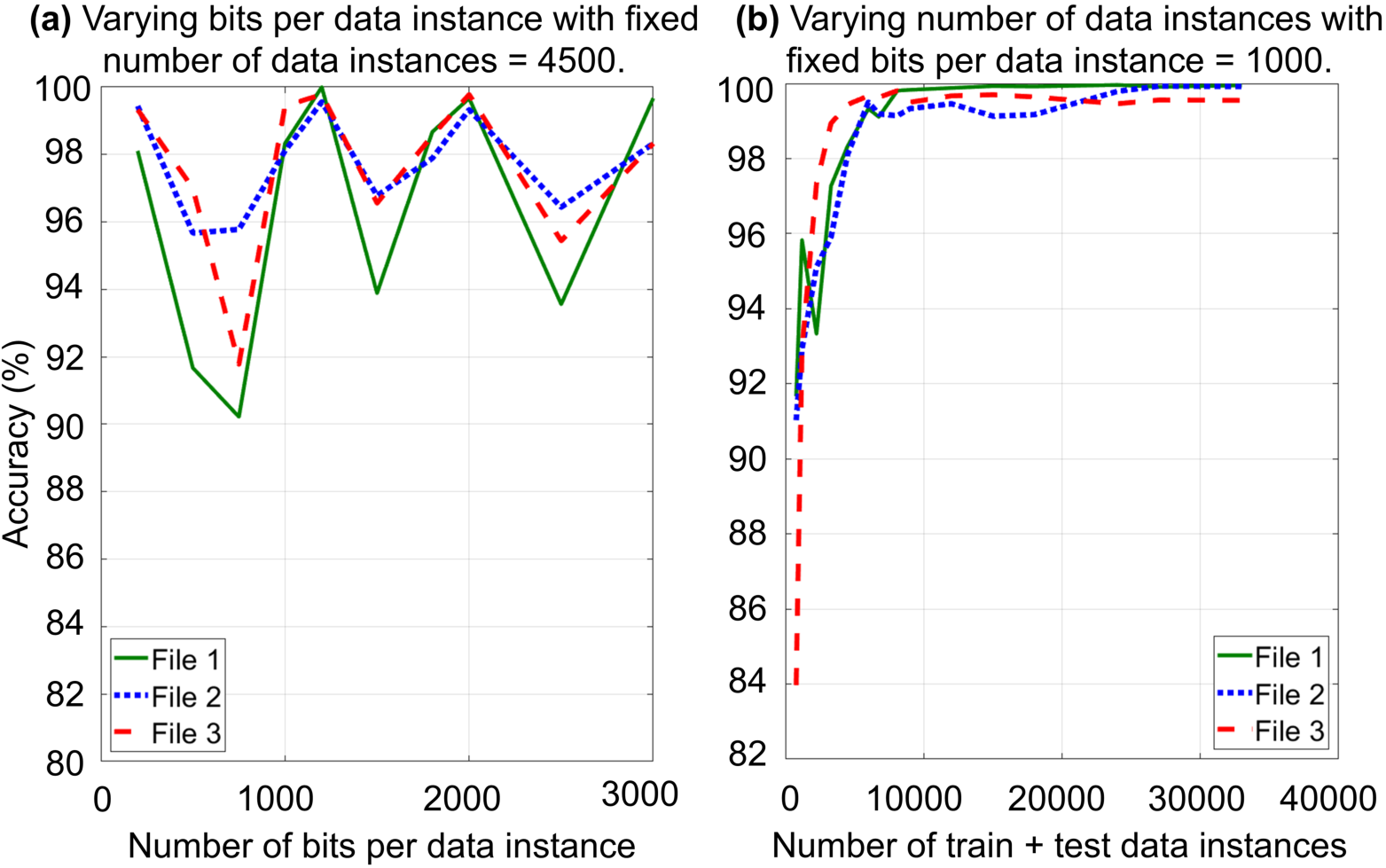} 
\centering
\vspace{-2mm}
\caption{Turbulence classification test accuracy scores for different files with 10 minute stabilization time after activation of the turbulence generators (data were taken from the beginning of each file).}
\label{fig:varybitsandlength10mins}
\vspace{-5mm}
\end{figure}

In  the central column (zero stabilization time) of Table~\ref{table:file123}, an obvious elevation in classification accuracy is noted as we progress temporally from file 1 to file 3. These results reveal a dynamic but converging impact of turbulence over time, leading to improved prediction accuracy. This trend is further illustrated in both subplots of Fig.~\ref{fig:varybitsandlength}, where classification accuracies associated with file 2 consistently surpass those linked to file 1. Moreover, the accuracies with file 3 exceed those with file 2 in the majority of cases. However, this effect is not evident in the results obtained with the data collected after 10 minutes of turbulence stabilization, as all three accuracy scores are nearly identical in the rightmost column of Table~\ref{table:file123}. Furthermore, in Fig.~\ref{fig:varybitsandlength10mins}(a), the accuracy scores with different files (with the same waiting period) are observed to vary randomly with all values consistently exceeding 90\%. This substantiates the earlier inference that the turbulence effect becomes stabilized over time, yielding similar and higher prediction accuracies across all three files.

In Fig.~\ref{fig:varybitsandlength}(a), the increase in the number of bits per data instance surprisingly introduces \emph{greater} ambiguity in the data due to the temporal variation in the effect of turbulence, leading to a decline in turbulence prediction accuracies. However, file 3, characterized by a comparatively stabilized turbulence level, exhibits a more gradual decrease in classification accuracy compared to the other two files. This suggests that when extra time is granted to allow the turbulence to stabilize, decreasing accuracy trends should not be observed.  This is immediately verified by the 10 min delayed data shown in Fig.~\ref{fig:varybitsandlength10mins}(a).

In Figs.~\ref{fig:varybitsandlength}(b) and~\ref{fig:varybitsandlength10mins}(b), the increase in the number of data instances, equivalent to introducing a longer time duration for the data considered for training and testing, was explored while maintaining a constant number of bits per data at 1000. As the number of data instances increases, turbulence classification accuracy initially rises for all three files, which is a common observation in most ML-based classifications~\cite{brain1999effect}. In Fig.~\ref{fig:varybitsandlength}(b), the accuracies reach a peak before gradually decreasing for files 1 and 2, but the accuracy nearly saturates after the initial increase for file 3. This behavior is again attributed to file 3's temporal position relative to the other two files.  For file 3, the turbulence has stabilized more, and consideration of more data instances (i.e. longer time duration) thereby enhances classification by accentuating the distinctions among turbulence levels. After 10~min, the turbulence has thoroughly stabilized, therefore all three files show a similar trend in Fig.~\ref{fig:varybitsandlength10mins}(b), saturating after reaching a knee point. Thus, the results presented in Figs.~\ref{fig:varybitsandlength} and~\ref{fig:varybitsandlength10mins} validate the time-dependent nature of turbulence in the optical channel and its very practical effect on ML-based turbulence classification.

The results also reveal some limitations of our method.  First, this method classifies discrete turbulence levels, but does not quantify $C_n^2$ or its variation along a path, which may be beneficial data in future practice. To that end, future work may incorporate the scintillation index $\sigma_I^2$ into the ML framework, since this can be measured without any additional hardware. Rytov or Kolmogorov theory could then be used to link $\sigma_I^2$ to $C_n^2$ \cite{andrews2005laser}. Indeed, this link may even be built into the ML framework. Second, the data reveal a time scale associated with the dynamic behavior of turbulence.  During transient changes in turbulence level, ML performed more poorly, however this could be avoided by a relatively short stabilization period on the order of 1~min or less.  The atmosphere regularly exhibits both short term turbulence fluctuations (tens of minutes) and long term fluctuations from morning to evening \cite{odintsov2019determination,fiorino2021re}.  It will be important in future ML endeavors to carefully regard all such time scales and a robust ML algorithm would account for this.  It is promising, however, that the algorithm's accuracy was capable of converging with only about 1~min of stabilization.  This suggests the ML method could be used for trend analysis to predict upcoming outages in real-time.  

The results also suggest the ML method may enhance the performance of real world systems. Existing measurement and modeling techniques capture turbulence variations in both vertical or horizontal $C_n^2$ profiles. However, low earth orbit satellites transition from horizon to overhead (to some extent) and back to horizon rapidly and during every orbital pass. Therefore, the beam may traverse all possible combinations of vertical and horizontal propagation through the atmosphere, a situation for which existing techniques are far less capable.  For example, it would be impractical to install sonar sounding stations with the required density to predict turbulence effects for all possible satellite links within range of a single ground station.  Since the ML technique utilizes only the optical data beam, the method demonstrated here is highly economical for coarsely monitoring turbulence in a large region surrounding the ground station. This could provide a ``communication weather map'' built up by the ground stations themselves without need for any additional infrastructure. This could also enable additional ML techniques on a broader scale to predict FSO outages, allowing the entire communication network to react by switching to RF bands or routing data to other FSO ground stations. These are the subjects of future investigations. 

Last, in real world FSO communications with fog, clouds and pointing errors, stable and known turbulence conditions do not exist for training the ML model. However, such conditions are only required during training. In the testing phase of this work, the levels of turbulence were not known in advance. Additionally, in practice, training could be restricted to clear weather days or using a few dedicated systems with more sophisticated and independent hardware for turbulence characterization. Trained models could then be deployed on simple FSO stations and still reap the benefits of turbulence prediction.

%\vspace{-4mm}
\section{Conclusion}
\label{sec:Conclusion}

This research presented a novel, empirically-validated approach to affirm the possibility of predicting discrete turbulence levels within an optical data channel by utilizing ML, specifically the XGBoost model.  The method consistently achieved an accuracy of $>98$\% so long as the dynamic changes in turbulence were allowed to stabilize over about 1~min.  A temporal analysis revealed that prediction accuracy could actually degrade with more training data during turbulence stabilization periods. The study also provides valuable insights into optimizing training strategies by examining the impact of the number of bits per data instance and the number of data instances on prediction accuracy. Looking ahead, future research directions include exploring ML models that better adapt to real-time dynamics of turbulence and provide more quantitative data about the channel.  Tradeoff studies considering the hardware constraints for practical ML deployment and training, may also be compared to existing methodologies.  By bridging the gap between ML and FSO communication, this research contributes to the development of more resilient and efficient communication systems.

\bibliographystyle{IEEEtran}
\bibliography{IEEEabrv,IEEEexample}

\end{document}